\documentclass[%
reprint,
 superscriptaddress,
 amsmath,amssymb,
 aps,
 prl,
]{revtex4-1}

\usepackage[usenames]{color}
\usepackage{graphicx}
\usepackage{dcolumn}
\usepackage{bm}

\begin{document}

\preprint{APS/123-QED}

\title{Measurement of pretzelosity asymmetry of charged pion production in
Semi-Inclusive Deep Inelastic Scattering on a polarized $^3$He target}



\author{Y.~Zhang} \email[Corresponding author: ]{yizhang@lzu.edu.cn}
\affiliation{Lanzhou University, Lanzhou 730000, Gansu, People's Republic of China}
\author{X.~Qian}
\affiliation{Duke University, Durham, NC 27708}
\affiliation{Physics Department, Brookhaven National Lab, Upton, NY 11973}
\author{K.~Allada}
\affiliation{University of Kentucky, Lexington, KY 40506}
\affiliation{Massachusetts Institute of Technology, Cambridge, MA 02139}
\author{C.~Dutta}
\affiliation{University of Kentucky, Lexington, KY 40506}
\author{J.~Huang}
\affiliation{Massachusetts Institute of Technology, Cambridge, MA 02139}
\affiliation{Los Alamos National Laboratory, Los Alamos, NM 87545}
\author{J.~Katich}
\affiliation{College of William and Mary, Williamsburg, VA 23187}
\author{Y.~Wang}
\affiliation{University of Illinois, Urbana-Champaign, IL 61801}
\author{K.~Aniol}
\affiliation{California State University, Los Angeles, Los Angeles, CA 90032}
\author{J.R.M.~Annand}
\affiliation{University of Glasgow, Glasgow G12 8QQ, Scotland, United Kingdom}
\author{T.~Averett}
\affiliation{College of William and Mary, Williamsburg, VA 23187}
% ...
\collaboration{The Jefferson Lab Hall A Collaboration}
\noaffiliation

\date{\today}

\begin{abstract}
An experiment to measure single-spin asymmetries in semi-inclusive production of charged
pions in deep-inelastic scattering on a transversely polarized $^3$He target was 
performed at Jefferson Lab in the kinematic region of $0.16<x<0.35$ and 
$1.4<Q^2<2.7$ ${\rm GeV^2}$. The pretzelosity asymmetries on $^3$He, which can be 
expressed as the convolution of the $h^\perp_{1T}$ transverse momentum dependent 
distribution functions and the Collins fragmentation functions in the leading order, 
were measured for the first time. Using the effective polarization approximation, we 
extracted the corresponding neutron asymmetries from the measured $^3$He asymmetries 
and cross-section ratios between the proton and $^3$He.
 Our results show that for both
$\pi^{\pm}$ on $^3$He and on the neutron the pretzelosity asymmetries are consistent
with zero within experimental uncertainties.
\end{abstract}
\pacs{Valid PACS appear here}

\maketitle



Studies of nucleon structure have been and still are at the frontier of understanding 
how quantum chromodynamics (QCD) works in the non-perturbative region. It has been known
for decades that the nucleon is composed of quarks and gluons, however how quarks and 
gluons contribute to the elementary properties of a nucleon from QCD is still an open 
question. Among these properties, the nucleon spin has been at the center of interest
for more than two decades since the original discovery by the European Muon 
Collaboration in 1988~\cite{EMC1988}, that quark spins were found to contribute only 
a small portion to the proton spin. 
In the last two decades, many polarized deep-inelastic scattering (DIS) experiments
~\cite{spin_review1} have been confirmed that the quark spin only contributes about 
25\% with significantly improved precision. In more recent years, efforts have also 
been devoted to the determination of the gluon spin contribution to the nucleon spin 
both from polarized DIS and from polarized proton-proton collision measurements
~\cite{spin_review2}. Recently, new results~\cite{RICH_spin_1,RICH_spin_2, QCD_SPIN}
from the RHIC-spin program suggest that the gluon spin may contribute to the proton 
spin only at a level comparable to that of quark spins. These findings suggest that
the orbital angular momentum (OAM) of the quarks and gluons, the most elusive piece,
may be the largest contributor to the nucleon spin.

In recent years, major theoretical and experimental efforts have focused on 
accessing OAM in the nucleon. The development of the general parton distribution 
functions (GPDs)~\cite{GPD_review} and the transverse-momentum-dependent 
parton distribution functions (TMDs)~\cite{TMD_review} provided not only 
three-dimensional imaging of the nucleon, but also promising ways to access OAM. By 
investigating the correlations between quark position and momentum, GPDs supply a new
way to characterize the contribution of the orbital motion of quarks to the spin of 
the nucleon. 
On the other hand, TMDs investigate the parton distributions in three-dimensional 
momentum space and provide information about the relationship between the quark 
momenta and the spin of either the nucleon or the quark. Since most TMDs are expected
to vanish in the absence of quark orbital motion, they supply important and
complementary ways to access the contribution of OAM to the spin of the proton.

Among the 8 leading-twist TMDs, there are only three that remain non-zero after an
integration over the parton transverse momentum~\cite{TMD_review}. They are the 
unpolarized parton distribution function (PDF) $f_1$, the longitudinally polarized 
PDF $g_1$ (helicity), and the transversely polarized PDF $h_1$ (transversity). $f_1$
has been extensively studied for several decades. $g_1$ is also relatively well 
understood by continuous efforts started in the 1970s~\cite{spin_review1}. 
For transversity, although less known than the former two, pioneering studies have 
been made in recent years, both theoretically and experimentally~\cite{h1_review}. 
One of the least known TMDs, $h^{\perp}_{1T}$, referred to as pretzelosity, has drawn 
significant attention recently~\cite{Miller03, Miller07, H2008b, H2010a, C2011} due to
its intuitive relation to the quark OAM. It is one of the eight leading-twist PDFs, 
with the odd chirality which leads to an important consequence that there are only
quark pretzelosity distributions, with no gluonic counterparts. 

In a class of relativistic quark models~\cite{H2010a, C2011}, pretzelosity can be 
expressed as the difference between the helicity and the transversity. This relation 
can be intuitively understood as that in a moving nucleon the difference in 
polarization of a quark in the longitudinal and transverse direction is due to the
fact that boosts and rotations do not commute. A non-zero value of the pretzelosity
is a direct consequence of this relativistic nature of quark motion.
Another interesting feature is that pretzelosity emerges from the 
interference of quark wave-function components with a difference of two units of orbital 
momentum~\cite{Burkardt2007}. Pretzelosity is the only leading-twist TMD with this 
unique feature. In quark models, the quark OAM can be directly accessed via 
pretzelosity~\cite{H2010a, C2011}. This finding was first obtained in a quark-diquark 
model~\cite{Jun2009} and a bag model~\cite{H2008b}, and confirmed later in a large 
class of quark models based on spherical symmetry~\cite{C2011}. 

Experimentally, pretzelosity is suppressed in the inclusive DIS processes due to its 
chiral-odd property. However, combined with another chiral-odd object such as the Collins 
fragmentation function~\cite{Collins1993}, it leads to a measurable effect in 
semi-inclusive DIS (SIDIS)~\cite{Feng2009} in which a leading hadron is detected in 
addition to the scattered lepton. Specifically, with an unpolarized lepton beam 
scattered from a transversely polarized nucleon target, a non-zero $h^{\perp}_{1T}$ 
would produce an azimuthal-angular dependent single-spin asymmetry (SSA) in the
differential cross sections of the scattered lepton and the leading hadron,
with respect to the target spin direction.

Following the Trento convention~\cite{Alessandro2004}, the azimuthal-angular dependence 
of the target SSA at the leading twist can be written as:
\begin{equation}
\begin{aligned}
 A_{UT}(\phi_h,\phi_s) =& \frac{1}{P} \cdot
\frac{Y(\phi_h,\phi_s)-Y(\phi_h,\phi_s+\pi)}{Y(\phi_h,\phi_s)+Y(\phi_h,\phi_s+\pi)} \\
\approx & A^{C} \cdot \sin(\phi_h+\phi_s) + A^{S} \cdot \sin(\phi_h-\phi_s) \\
& + A^{p}\cdot \sin(3\phi_h-\phi_s),
\end{aligned}
\label{asymmetry_formula}
\end{equation}
where the subscript $U$ stands for the unpolarized beam, $T$ stands for the
transversely polarized target,
$P$ is the polarization of the target, $Y$ is the normalized yield, $\phi_h$ is the
angle between the lepton plane and the hadron plane, which is defined by the hadron 
momentum direction and the virtual photon momentum direction, and $\phi_s$ is the angle
between the target spin direction and the lepton plane. The higher-twist terms have 
been neglected. The three leading-twist asymmetries~\cite{Anselmino2011} correspond to
the Collins asymmetry ($A^C$), the Sivers asymmetry ($A^S$) and the pretzelosity
asymmetry ($A^p$). The Collins asymmetry is due to the transversity distribution 
function convoluted with the Collins fragmentation function, while the Sivers asymmetry 
is the Sivers distribution function convoluted with the unpolarized fragmentation 
function. The last term, referred to as the pretzelosity asymmetry, is the pretzelosity
distribution function convoluted with the Collins fragmentation function. As shown in 
Eq.~(\ref{asymmetry_formula}), these three terms have different azimuthal angular 
dependences, therefore it is possible to separately determine each term by fitting the
angular dependence.

The HERMES collaboration carried out the first measurement of the Collins and Sivers
asymmetries~\cite{hermes_sidis_1} with electron and positron beams on a transversely 
polarized proton target. The COMPASS collaboration performed measurements with a 
muon beam on transversely polarized proton~\cite{compass_sidis_1} and deuteron 
targets~\cite{compass_sidis_2}. In Hall A at Jefferson Lab (JLab), an exploratory 
experiment E06-010~\cite{xqian_prl, jin_prl} was carried out, for the first time using
an electron beam on a transversely polarized $^3$He target. The extracted Collins 
and Sivers asymmetries were published recently~\cite{xqian_prl}. In extracting 
these asymmetries, the pretzelosity term was not included. The uncertainty due to 
this treatment was estimated and included in the systematic uncertainty.

\begin{figure}
 \begin{center}
  \includegraphics[width=.3\textwidth]{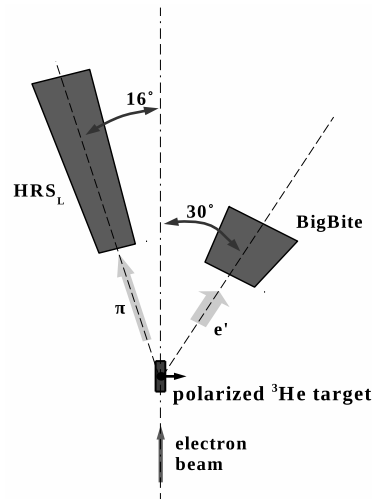}
  \caption{The schematic top view of the experiment E06-010.}
  \label{layout}
 \end{center}
\end{figure}

In this letter, we present the results of the pretzelosity asymmetry extracted from the
JLab E06-010 data. As shown in Fig.~\ref{layout}, in the experiment a 5.9-GeV electron 
beam was incident on a transversely polarized gaseous $^3$He target with an average 
current of 12 $\mu$A. The target~\cite{halla_nim} was polarized by spin-exchange 
optical pumping~\cite{Walker1997} of a Rb/K mixture. The scattered electrons were 
detected using the BigBite spectrometer~\cite{halla_nim} at beam right with a 
solid-angle acceptance of $\sim$64 msr. Three sets of drift chambers with eighteen wire
planes in total were used for tracking.
Lead-glass pre-shower and shower detectors were used to identify electrons. The hadron
contamination in the electron sample was suppressed to below 2\% in the kinematic
range of 0.6-2.5 GeV/$c$. The produced hadrons were detected in the left arm of the high
resolution spectrometers~\cite{halla_nim} (${\rm LHRS}$) at beam left. A gas Cherenkov
detector and two layers of lead-glass detectors provided a clean separation of 
electrons from pions. An aerogel Cherenkov detector and the coincident time-of-flight
technique were employed to distinguish pions from kaons and protons. The path
length of about 25 meter from the target to the ${\rm LHRS}$ focal plane made it
possible to make an accurate time-of-flight measurement.

\begin{figure}
 \begin{center}
  \includegraphics[width=.49\textwidth]{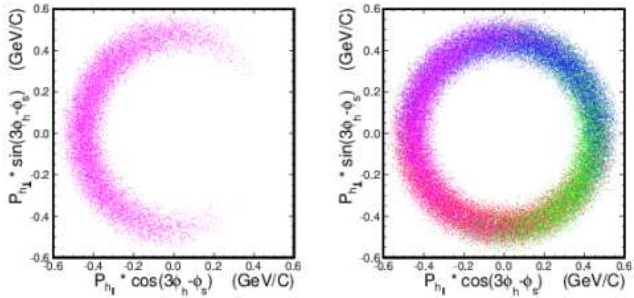}
  \caption{Phase space coverage of the data in the lowest {\it x}-bin modulated by 
  the ${\rm \sin(3\phi_h - \phi_s)}$ term. The left panel shows the data in only one 
  target spin orientation, while the right panel shows the data in all four orientations.}
  \label{PhaseSpace}
 \end{center}
\end{figure}

To extract moments of the SSA, it is important to have the data cover the range of 
azimuthal angle in phase space as full as possible. In the case of pretzelosity 
asymmetry, the azimuthal angle is $(3\phi_h-\phi_s)$ in a range of $[0,2\pi]$.
In the experiment, the BigBite and the left HRS spectrometer covered only part of the
$2\pi$ angular range. To increase the kinematic coverage, four different 
target spin orientations orthogonal to the beam direction, transverse left, transverse 
right, vertical up and vertical down, were used. For each target spin orientation the 
spectrometers covered only a section of the phase space as shown in the left panel 
of Fig.~\ref{PhaseSpace} (target spin vertical up). However, data from all four 
orientations, when combined, covered the full angular range as shown in the right panel
of Fig.~\ref{PhaseSpace}, where magenta, green, red and blue are for horizontal beam 
left, horizontal beam right, vertical up and vertical down, respectively. In order to
achieve target polarizations in these four orientations, three pairs of mutually
orthogonal Helmholtz coils were employed. During the experiment, the target spin
direction was flipped every twenty minutes using the adiabatic fast passage technique,
while the magnetic holding field direction and strength remained unchanged. 

Several kinematic cuts were used to select SIDIS events: the square of the
four-momentum transfer $Q^2 > 1$ GeV$^2$, the invariant mass of the virtual 
photon-nucleon system $W > 2.3$ GeV and the invariant mass of the undetected final
state particles $W' > 1.6$ GeV. Data were divided into 4 Bjorken-$x$ bins with approx
equal statistics. The central kinematics are presented in Table~\ref{tab1}.
\begin{table} [!ht]
\caption{Central kinematics for the four $x$ bins. The fractional e$^-$ energy loss $y$,
the hadron energy fraction $z$ with respect of electron energy transfer, and the 
transverse momentum $P_{h\perp}$ are all defined following the notation
in Ref.~\cite{Alessandro2004}.}
\label{tab1}
\begin{tabular}[b]{ c c c c c c c }
\hline \\
$x$ & $Q^2$ GeV$^2$ & $y$ & $z$ & $P_{h\perp}$ GeV/c & $W$ GeV & $W'$ GeV \\
\hline 
0.156 & 1.38 & 0.81 & 0.50 & 0.44 & 2.91 & 2.07 \\
0.206 & 1.76 & 0.78 & 0.52 & 0.38 & 2.77 & 1.97 \\
0.265 & 2.16 & 0.75 & 0.54 & 0.32 & 2.63 & 1.84 \\
0.349 & 2.68 & 0.70 & 0.58 & 0.24 & 2.43 & 1.68 \\
\hline 
\end{tabular}
\end{table}
To minimize the systematic uncertainties, the data taken between each of the two flips 
of the target spin were further divided in two equivalent sections. Two following data
sets with opposite spin directions formed a local pair, from which a local
raw asymmetry was extracted. Throughout the experiment, $\sim$2850 such local raw
asymmetries were combined to form the total raw asymmetry. Pretzelosity
moments were extracted from the total raw asymmetry by a fit to 
Eq.~(\ref{asymmetry_formula}) in two-dimensional $(\phi_h, \phi_s)$ bins. 

In the polarized $^3$He target, a small amount ($\sim1\%$ in volume) of N$_2$ gas 
was mixed with $^3$He to reduce depolarization effects. The nitrogen nuclei also
contributed to the total measured yield and diluted the measured raw asymmetries. To 
obtain the asymmetries on $^3$He, a correction for the nitrogen dilution was applied to
the raw asymmetries, as shown in Eq.~(\ref{nitrogen_dilution})
\begin{equation}
	A^{P}_{\rm ^3He} = A^{P}_{raw} / \left( 1 -
	\frac{N_{\rm N_2}\sigma_{\rm N_2}}{N_{\rm N_2}\sigma_{\rm N_2} + 
		N_{\rm ^3He}\sigma_{\rm ^3He}} \right).
\label{nitrogen_dilution}
\end{equation}
In Eq.~(\ref{nitrogen_dilution}) the $\sigma$'s are the unpolarized cross sections and
the $N$'s are the number densities. In the experiment, the cross section ratio
$\sigma_{^3He}/\sigma_{N_2}$ was measured through dedicated data taking with a 
reference target cell filled with a known amount of unpolarized $^3$He and N$_2$ gases.
The number densities of $^3$He and N$_2$ in the polarized $^3$He target were verified
by taking elastic scattering data on both the reference target and the production $^3$He 
target~\cite{yzhang}. Another significant systematic correction was due to background
electrons in the SIDIS electron samples, mainly due to pair production. This is 
especially significant in the lowest $x$ bin. Dedicated data were taken with the BigBite
spectrometer in reversed polarity to measure the yield of the coincident 
$(e, e^{+\prime}\pi^{\pm})$ events, which is identical to the yield of electrons from
charge-symmetric pair production. This effect was corrected as a dilution since the
measured asymmetries of the coincident $(e, e^{+\prime}\pi^{\pm})$ events were
consistent with zero.

In the analysis, the systematic fitting uncertainties due to omission of the other
$\phi_h$- and $\phi_s$-dependent terms were taken into account, including the 
Cahn ($\langle \cos(\phi_h)\rangle$) and Boer-Mulders ($\langle \cos(2\phi_h)\rangle$) 
effects, higher-twist terms ($\langle \sin(\phi_s) \rangle$ and $\langle 
\sin(2\phi_h-\phi_s)\rangle$) and the $A_{UL}$ terms ($\langle \sin(\phi_h)\rangle$ 
and $\langle \sin(2\phi_h)\rangle$)~\cite{Bacchetta2007, Anselmino2011}. 
In particular, the $A_{UL}$ terms were induced by a small longitudinal component of the 
target polarization in the center-of-mass frame of the SIDIS process. Of all these 
effects, the $\langle \sin(2\phi_h-\phi_s)\rangle$ term was largest ($\sim16\%$ of the
statistical uncertainty), followed by the $\langle \sin(\phi_s)\rangle$ term 
($\sim14\%$ of the statistical uncertainty). To take into account the systematic 
uncertainty induced by $K^{\pm}$ contamination in $\pi^{\pm}$ example, the 
coincident $(e, e^{\prime}K^{\pm})$ events were selected and the $\sin(3\phi_h-\phi_s)$
term of the asymmetry was extracted by the maximum likelihood method. Then, the 
systematic uncertainty was evaluated as the difference between the 
$\sin(3\phi_h-\phi_s)$ terms of the $(e, e^{\prime}\pi^{\pm})$ and the 
$(e, e^{\prime}K^{\pm})$ samples, weighted by the contamination ratios of the 
$K^{\pm}$ in $\pi^{\pm}$ samples. Other ingredients of the systematic uncertainties 
included the yield drift, the target polarization, the target-density fluctuation, the
detector tracking efficiency, the live time, the nitrogen dilution, and the photon
contamination in the BigBite spectrometer. Since those ingredients have no azimuthal
angular dependence and share the same data set of \cite{xqian_prl}, they have the same
uncertainties as in \cite{xqian_prl}.

\begin{figure}
 \begin{center}
  \includegraphics[width=.49\textwidth]{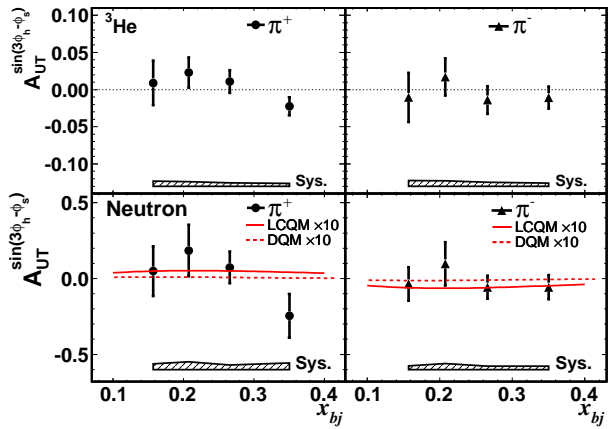}
  \caption{The extracted pretzelosity asymmetries on $^3$He nuclei (top panels) and 
on the neutron (bottom panels) are shown together with uncertainty bands for both
$\pi^+$ and $\pi^-$ electron production.}
  \label{result_plot}
 \end{center}
\end{figure}

The extracted $h^{\perp}_{1T}$ results for the $^3$He target are shown in the top 
two panels of Fig.~\ref{result_plot}. Only the statistical uncertainties are included 
in the error bars.  The experimental systematic uncertainties are combined in quadrature
and shown as the band labeled ``Sys.''. All the extracted $\pi^+$ and $\pi^-$ 
pretzelosity terms are small and consistent with zero within the uncertainties, further
supporting the assumption in previous analysis~\cite{xqian_prl} that the inclusion
of pretzelosity term has little effect on extraction of the Collins and Sivers term.

To extract the pretzelosity asymmetries on the neutron, the effective polarization
method was used:
\begin{equation}
  A^{P}_{n} = \frac{1}{(1-f_p)P_n} \left(A^{P}_{he} - f_p A^{P}_p P_p\right),
\label{proton_dilution}
\end{equation}
where the proton dilution factor $f_p \equiv 2 \sigma_p / \sigma_{^3 {\rm He}}$ was
obtained by measuring the yields of unpolarized proton and unpolarized $^3$He
targets at the same kinematics. The same model uncertainty due to final-state
interactions as in \cite{xqian_prl} was taken into account for $f_p$.
$P^n = 0.86^{+0.036}_{-0.02}$ and $P^p = -0.028^{+0.009}_{-0.004}$ are the effective
polarizations of the neutron and proton in a $^3$He nucleus~\cite{Xiaochao2004,Ethier2013}. 
Due to the scarcity of available data and the small effective polarization of the proton,
in this analysis no correction was applied to account for the effect due to the proton
asymmetry. The uncertainty due to this omission was estimated and included in the
systematic uncertainty. For positive pions at the highest $x$ bin, the asymmetry is 
magnified by nearly one order of magnitude from $^3$He to the neutron, due to the large
proton dilution.

The extracted pretzelosity moment on the neutron is shown in the bottom two panels of
Fig.~\ref{result_plot} and is compared with the quark-diquark model~\cite{Jun2009} and
light-cone constitute-quark model~\cite{Pass2008, Pass2009} calculations. Like in the 
two upper panels, the error bars shown only represent the statistical uncertainties,
while the bands labeled ``Sys.'' represent the systematic uncertainties.
Since the differences between the two model predictions are hardly visible compared to
the statistical uncertainties, the curves in the two panels are multiplied by a factor
of 10. The extracted neutron asymmetries of both $(e, e^{\prime}\pi^+)$ and $(e, 
e^{\prime}\pi^-)$ are consistent with zero. Compared to the $\sin(\phi_h+\phi_s)$ terms,
the $\sin(3\phi_h-\phi_s)$ terms are suppressed by a factor of order $k_\perp^2/M^2$
~\cite{Anselmino2011}, in which $k_\perp$ is the parton transverse momentum and $M$ is
the mass of the nucleon. As suggested in~\cite{Jun2009}, a large $P_{h\perp}$ coverage
such as that planned for future experiments~\cite{solid_white} with a higher statistical
precision, is required to observe any non-zero pretzelosity asymmetry. It is worth
mentioning that the small value for the asymmetry predicted by the quark-diquark model
(of the order of $10^{-3}$) is mainly due to kinematic suppression and hence does not
necessarily predict that $h_{1T}^{\perp}$ is small. In that calculation $h_{1T}^{\perp}$
is proportional to the OAM of the quarks, originating from a Melosh rotation of the
quark spin distribution between the instant and the light-cone frame, and thus can be a
significant contribution to the spin of the nucleon.


In summary, we present the first measurement of pretzelosity asymmetries on a 
transversely polarized $^3$He target, utilizing charged pion production in SIDIS
process. The asymmetries are consistent with zero within experimental uncertainties in
this kinematic region, and are also consistent with model expectations. This work 
demonstrates an experimental approach for studying the $h^{\perp}_{1T}$ TMD and lays a 
foundation for future high-precision measurements~\cite{solid_white}.


We acknowledge the outstanding support of the JLab Hall A staff and
Accelerator Division in accomplishing this experiment. This work was supported in part
by the U.S. National Science Foundation, and by U.S. DOE contract DE-AC05-06OR23177,
under which Jefferson Science Associates, LLC operates the Thomas Jefferson National
Accelerator Facility. This work was also supported by the National Natural Science 
Foundation of China under Grant No. 11135002 and No. 11120101004.


\bibliographystyle{model1a-num-names}
\bibliography{mylibrary}

\end{document}